\documentclass[11pt]{article}
\usepackage{fullpage,citesort,epsfig,psfrag}
\usepackage[normal,small]{caption}
\newcommand{\beq}{\begin{equation}}
\newcommand{\eeq}{\end{equation}}
\newcommand{\bea}{\begin{eqnarray}}
\newcommand{\eea}{\end{eqnarray}}

\def\Tr{{\rm Tr}}
\def\to{\rightarrow}
\newcommand{\be}{\begin{equation}}
\newcommand{\ee}{\end{equation}}
\newcommand{\bq}{\begin{eqnarray}}
\newcommand{\eq}{\end{eqnarray}}

\def\ie{{\it i.e.\ }}
\def\m@th{\mathsurround=0pt }
\def\leftrightarrowfill{$\m@th \mathord\leftarrow \mkern-6mu \cleaders\hbox{$\mkern-2mu \mathord- \mkern-2mu$}\hfill
 \mkern-6mu \mathord\rightarrow$}
\def\overleftrightarrow#1{\vbox{\ialign{##\crcr
     \leftrightarrowfill\crcr\noalign{\kern-1pt\nointerlineskip}
     $\hfil\displaystyle{#1}\hfil$\crcr}}}

\begin{document}
\setlength{\captionmargin}{20pt}

\renewcommand{\thefootnote}{\fnsymbol{footnote}}
\begin{titlepage}
\begin{flushright}
UFIFT-HEP-02-9\\
hep-th/0203167
\end{flushright}

\vskip 3cm

\begin{center}
\begin{Large}
{\bf A Worldsheet Description of Planar Yang-Mills Theory\footnote{
This work was supported in part by the Monell Foundation
and in part by the Department
of Energy under Grant No. DE-FG02-97ER-41029. 
}}
\end{Large}

\vskip 2cm
{\large 
 Charles B. Thorn\footnote{E-mail  address: {\tt thorn@phys.ufl.edu}}
}

\vskip 0.5cm
{\it School of Natural Science, Institute for Advanced Study,
Princeton NJ 08540}
\vskip0.20cm 
and
\vskip0.20cm
{\it Institute for Fundamental Theory\\
Department of Physics, University of Florida,
Gainesville FL 32611}


\vskip 1.0cm
\end{center}

\begin{abstract}\noindent
We extend previous work by developing
a worldsheet description of non-abelian
gauge theory (Yang-Mills). This task requires the
introduction of Grassmann variables on the world sheet
analogous to those of the Neveu-Schwarz-Ramond formulation
of string theory. A highlight of our construction is that
once the three gluon vertices of Yang-Mills Feynman
diagrams are given a worldsheet description, the worldsheet formalism
automatically produces all of the quartic vertices.
\end{abstract}
\vfill
\end{titlepage}
\section{Introduction}
\label{chap1}
Several months ago, Bardakci and I proposed \cite{bardakcit} 
a way of expressing the planar Feynman diagrams, selected
by 't Hooft's $N_c\to\infty$ limit \cite{thooftlargen}, 
of a quantum matrix field theory in the
language of light-cone interacting string diagrams \cite{mandelstam,gilest}.
As a paradigm for our method we treated the simplest example
of $\Tr\phi^3$ theory, but stressed that our methods should
be applicable to a wide range of quantum field theories.

In this article we extend the program to include
non-abelian gauge theories (Yang-Mills theories).
The extension involves several new features. 
The light-cone gauge ($A_-=0$)
Feynman rules for gauge theories involve both
cubic and quartic vertices. The cubic vertices are
linear in the transverse momenta entering the vertex,
and also depend on the spin of the gauge bosons.
In addition, the contribution to the
cubic vertex that arises from the elimination of the
longitudinal potential $A_+$ is a rational function of the
$p^+_i$ entering the vertex.
The quartic vertices don't depend on the transverse
momentum, but they are spin dependent and the induced
``Coulomb exchange'' quartic is a rational function of the
$p^+_i$ entering the vertex. Thus the worldsheet description
must describe, in addition to the flow of momentum through
a complicated planar graph, the flow of gluon spin through the
graph. Further, the cubic vertices display a spin-orbit
coupling.

We shall introduce vector-valued Grassmann variables
${\bf S}_{1,2}(\sigma,\tau)$ of the Neveu-Schwarz-Ramond type to
keep track of the spin flow in a large diagram. 
In addition to contributing to the bulk worldsheet
action, these variables appear in insertion factors at
the interaction points of the worldsheet. These
insertion factors provide the necessary momentum and spin dependence
of the cubic vertices. We shall also need to supplement the
ghost system introduced in \cite{bardakcit} to locally
describe the new factors of $p^+$ that occur in the vertices.
We use an $x^+,p^+$ lattice to give a concrete
meaning to our construction, especially at interaction points
on the world sheet. But in the bulk of the world sheet,
away from boundaries, the worldsheet action has a rather
simple formal continuum limit
\bea
S=\int d\tau d\sigma\ \left[{{\bf q}^{\prime2}\over2}
+{\bar{\bf S}}_2\cdot{\bf S}^\prime_2 
-{\bar{\bf S}}_1\cdot{\bf S}^\prime_1+{\rm Ghost~Terms}\right] .
\eea
Here the prime denotes $\partial/\partial\sigma$. The bosonic
worldsheet variable ${\bf q}$ was introduced in \cite{bardakcit}.
Its derivative  is the density of transverse momentum
on a bit of world sheet: that is ${\bf q}^\prime d\sigma$
is the transverse momentum carried by the element $d\sigma$.
The noteworthy feature here is that there are no time
derivatives in the bulk: all dynamics occurs at the
boundaries of the worldsheet.

For the quartic vertices of the usual Feynman
rules, the worldsheet formalism provides
a pleasant surprise. Initially it seemed evident that a quartic
vertex would introduce nonlocal elements into the worldsheet
description. This is because four worldsheet strips of
varying width, corresponding to the different $p_i^+$ carried by
each external leg into a vertex, can't be glued together
at a point. It is therefore gratifying that these
quartic vertices are produced from pairs of cubic
vertices by the worldsheet formalism. This comes about as
follows. The worldsheet description of the cubic vertex,
which is linear in the transverse momentum, requires the
insertion of a factor of ${\bf q}^\prime(\sigma_1,\tau_1)$
for $\sigma_1,\tau_1$ near the interaction point. When a
four point amplitude is built from cubics there are
two such insertions at $\sigma_1,\tau_1$ and $\sigma_2,\tau_2$,
say. When $\tau_1\neq\tau_2$ this double insertion just provides
the momentum factors for the two cubic vertices. However,
for $\tau_1=\tau_2$ the double insertion provides an additional
``quantum fluctuation'' term whose value precisely gives
the quartic vertex. Since the quartic vertices are taken
care of by the locally specified worldsheet formalism,
no worldsheet nonlocal elements need be introduced.

In Section 2 we review the worldsheet
construction for the scalar theory given in Ref.~\cite{bardakcit},
making some minor adjustments that will be helpful for the 
Yang-Mills construction.
Section 3 contains the main result of the paper,
the extension of the worldsheet formalism to Yang-Mills theories. 
Section 4 contains concluding remarks.

\section{Review of Scalar Theory${}^\dag$}
\footnotetext{The new features in this review of 
Ref.~\cite{bardakcit} were developed in collaboration with K. Bardakci.}
\label{sec2}

\subsection{Propagator}
\label{sec2.1}
We use light-cone coordinates defined
as $x^\pm=(x^0\pm x^3)/\sqrt2$. The remaining components
of $x^\mu$ will be distinguished by Latin indices, or
as a vector by bold-face type. The components of any four vector
$v^\mu$ will be $(v^+,v^-,{\bf v})$ or $(v^+,v^-,v^k)$.
The Lorentz invariant
scalar product of two four vectors $v,w$ is written
$v\cdot w={\bf v}\cdot{\bf w}-v^+w^--v^-w^+$. We shall
select $x^+$ to be our quantum evolution parameter, and we
recall that the Hamiltonian conjugate to this time is
$p^-$. A massless on-shell particle thus has the ``energy''
$p^-={\bf p}^2/2p^+$.

Central to the worldsheet construction of \cite{bardakcit}
is the mixed representation ($x^+,p^+,{\bf p}$) of the propagator 
of the free massless scalar field \cite{thooftlargen}:
\begin{eqnarray}
\Delta({\bf p},p^+,x^+)
\equiv\int {dp^-\over2\pi i}e^{-ix^+p^-}{1\over p^2-i\epsilon}
={\theta(x^+)\over2p^+}e^{-ix^+{\bf p}^2/2p^+}\to
{\theta(\tau)\over2p^+}e^{-\tau{\bf p}^2/2p^+},
\label{propagator}
\end{eqnarray}
where we assume $p^+>0$. Since we consider only planar
diagrams in this paper, we may suppress explicit
reference to color indices. Note that we have defined imaginary
time $\tau=ix^+$. We use imaginary $x^+$ 
for convenience to make integrals damped
instead of rapidly oscillating.

We set up a lattice worldsheet \cite{gilest} by discretizing $\tau$
and $p^+$:
\begin{eqnarray}
\tau= ka,\qquad p^+=lm,\qquad {\rm for}\quad k,l=1,2,3,\ldots
\end{eqnarray}
Then the scalar propagator becomes
\begin{eqnarray}
\Delta\to{\theta(k)\over2lm}e^{-k(a/m){\bf p}^2/2l}
\end{eqnarray}
According to the traditional Feynman rules each propagator momentum
is integrated with measure $dp^+d{\bf p}/(2\pi)^3$.  Combining the
$1/2p^+$ factor with this measure produces
\bea
\int{dp^+\over2p^+}\int {d{\bf p}\over(2\pi)^3}\to
\sum_l\int{d{\bf p}\over2l(2\pi)^3}.
\eea
In our construction we associate the factor $1/2l(2\pi)^3$ with one
or both of the two vertices to which the line is attached. We
shall distribute the purely numerical factors symmetrically
between the two vertices, an $n$-leg vertex receiving a factor
$1/(16\pi^3)^{n/2}$. Each
vertex is also integrated with measure 
$$\int d\tau (2\pi)^3\delta\left(\sum p^+_i\right)
\delta\left(\sum{\bf p}_i\right)\to \sum_k{(2\pi)^3a\over m}\delta_{l_1+l_2+l_3,0}
\delta\left(\sum{\bf p}_i\right).$$
We shall suppress the delta functions but include the numerical factors
in the vertex. Thus all together each $n$-leg vertex is multiplied
by the numerical factors $a/2m(16\pi^3)^{(n-2)/2}$.
For cubics this factor is just $a/8m\pi^{3/2}$ and for quartics
it is $a/32m\pi^3$.

In contrast, we assign the factor $1/l$ asymmetrically to
only one of the two vertices attached to the propagator.
Since in light-cone quantization 
all propagation is forward in time, we can consistently
assign the factor to the {\it earlier}
of the two vertices connected by the line. Then a {\it fission}
vertex, in which one particle with $p^+=lm$ ``decays'' to two particles
with $p^+_1=l_1m$, $p^+_2=l_2m$ will be assigned the
factor $1/l_1l_2$, whereas a {\it fusion} vertex, which
is the time-reversal of the fission vertex, will be assigned
the factor $1/l$. 
All propagators are then simple 
exponentials $e^{-ka{\bf p}^2/2lm}$.

Consider a line carrying $M$ units of $p^+$. As in \cite{thooftlargen}
we write the total momentum as a difference  
${\bf p}\equiv {\bf q}_M-{\bf q}_0$. 
We also must introduce ghost variables $b,c$ on each
lattice site to provide crucial factors of $M$ when we
represent a single gluon as $M$ bits. We make use of 
the ghost integrals
\begin{eqnarray}
\int \prod_{i=1}^{M-1} {dc_idb_i\over2\pi} \exp\left\{{a\over m}
\left[b_1c_1+b_{M-1}c_{M-1}+\sum_{i=1}^{M-2}(b_{i+1}-b_i)(c_{i+1}-c_i)\right]
\right\}
&=&M\left({a\over2\pi m}\right)^{M-1}\; .
\label{ghostbits}
\end{eqnarray}
For simplicity of presentation, we restrict attention to
the case $d=2$, four dimensional space-time. Then we
only require one $b,c$ set of ghosts.
Thus for each time slice introduce the 
action
\bea
S&=&S_g+S_q\\
S_q&=&{a\over2m}\sum_j\sum_{i=0}^{M-1}({\bf q}^j_{i+1}-{\bf q}^j_i)^2\\
S_g&=&-{a\over m}\sum_j\left[b^j_1c^j_1+b^j_{M-1}c^j_{M-1}
+\sum_{i=1}^{M-2}(b^j_{i+1}-b^j_i)(c^j_{i+1}-c^j_i)\right]
\eea
 Then
\begin{eqnarray}
\exp\left\{-N{a\over m}
{({\bf q}_M-{\bf q}_0)^2\over2M}\right\}&=&
\int\prod_{j=1}^N\prod_{i=1}^{M-1} {dc^j_idb^j_i\over2\pi} 
d{\bf q}^j_i\ e^{-S_g-S_q}\equiv\int DcDbD{\bf q}\ e^{-S}.
\label{nstepbits}
\end{eqnarray}

In (\ref{nstepbits}) we have imposed Dirichlet boundary conditions
${\bf q}_{0,j}={\bf q}_0$ and ${\bf q}_{M,j}={\bf q}_M$. However,
to dynamically 
implement momentum conservation in the
path integral we integrate over  $q^j_M$ independently,
retain the Dirichlet boundary condition at $i=0$, where we impose ${\bf q}_0^j={\bf q}_0$, but insert
momentum conserving delta functions at the other end $i=M$.
It is convenient but not necessary to set ${\bf q}_0=0$.
For formal uniformity we also introduce extra ghost variables
$b_M^j,c_M^j$ which we integrate over with weight 
$\exp{2\pi \sum_j b^j_Mc^j_M}$, which just amounts to inserting a factor
of unity in the path integral.
The path integral for the free propagator is then given for $d=2$ by
\begin{eqnarray}
T_{fi}&=&
\int\prod_{j=1}^N\prod_{i=1}^{M} {dc^j_idb^j_i\over2\pi} 
d{\bf q}^j_i\prod_{j=0}^N{d{\bf y}_M^j\over(2\pi)^2}
\exp\left\{-S+2\pi \sum_j b^j_Mc^j_M-i\sum_{j=0}^N{\bf y}_M^j\cdot({\bf q}_M^{j+1}-{\bf q}_M^j)\right\}
\label{discretefreepi}
\end{eqnarray}
When
we draw discretized worldsheet diagrams (see Fig.~\ref{oneloopws}),
with time flowing up,
we shall distinguish the bulk and boundary degrees of freedom
by dotted and solid vertical lines respectively. 

\subsection{Cubic Interactions}
In this subsection we review the world sheet construction of the cubic
vertices in the planar
diagrams of $g\Tr\phi^3/3\sqrt{N_c}$ theory.
A fission vertex describes a
field quantum with momenta ${\bf Q},Mm$ transforming to
two field quanta with momenta ${\bf p},lm; ({\bf Q}-{\bf p}),(M-l)m$. 
Before the interaction
we have a single propagator treated as in Section~\ref{sec2.1}. 
After the interaction
we have two propagators. 
We assign the factor $ga/8l(M-l)m\pi^{3/2}$ to the vertex.
The factors $1/l(M-l)$ must be incorporated by a
worldsheet local treatment of the fission point.

This is done by altering the ghost integral near
the interaction point to force the ghost integral
to give unity instead of a factor of $M$ or $l(M-l)$.
For example, deleting the $i=M-1$ (or $i=0$) term in
the exponential of (\ref{ghostbits}) has precisely this effect:
\begin{eqnarray}
\int \prod_{i=1}^{M-1} {dc_idb_i\over2\pi} 
\exp\left\{{a\over m}
\left[b_1c_1+\sum_{i=1}^{M-2}(b_{i+1}-b_i)(c_{i+1}-c_i)\right]\right\}
&=&\left({a\over2\pi m}\right)^{M-1}\label{ghostunity1}\\
\int \prod_{i=1}^{M-1} {dc_idb_i\over2\pi} 
\exp\left\{{a\over m}
\left[b_{M-1}c_{M-1}+\sum_{i=1}^{M-2}(b_{i+1}-b_i)(c_{i+1}-c_i)\right]\right\}
&=&\left({a\over2\pi m}\right)^{M-1}
\label{ghostunity2}
\end{eqnarray}
Note, by the way, that if both the first and 
last terms are deleted the result is
zero:
\begin{eqnarray}
\int \prod_{i=1}^{M-1} {dc_idb_i\over2\pi} \exp\left\{{a\over m}
\sum_{i=1}^{M-2}(b_{i+1}-b_i)(c_{i+1}-c_i)\right\}
&=&0.
\label{ghostbit0}
\end{eqnarray}
This fact will be important in the worldsheet construction
for Yang-Mills theory.
We also get unity if we delete some interior ghost coupling term
labeled by $k$ 
\bea
\int \prod_{i=1}^{M-1} {dc_idb_i\over2\pi} 
\exp\left\{{a\over m}
\left[b_1c_1+b_{M-1}c_{M-1}+
\sum_{i=1 \atop i\neq k}^{M-2}(b_{i+1}-b_i)(c_{i+1}-c_i)\right]\right\}
&=&\left({a\over2\pi m}\right)^{M-1}\hskip-14pt.\hskip18pt
\eea

\begin{figure}[ht]
\vskip.5cm
\psfrag{'k'}{$k$}
\psfrag{'k+l'}{$k+l$}
\psfrag{'M'}{$M$}\psfrag{'M1'}{$M_1$}
\psfrag{'N'}{$N$}
\centerline{\epsfig{file=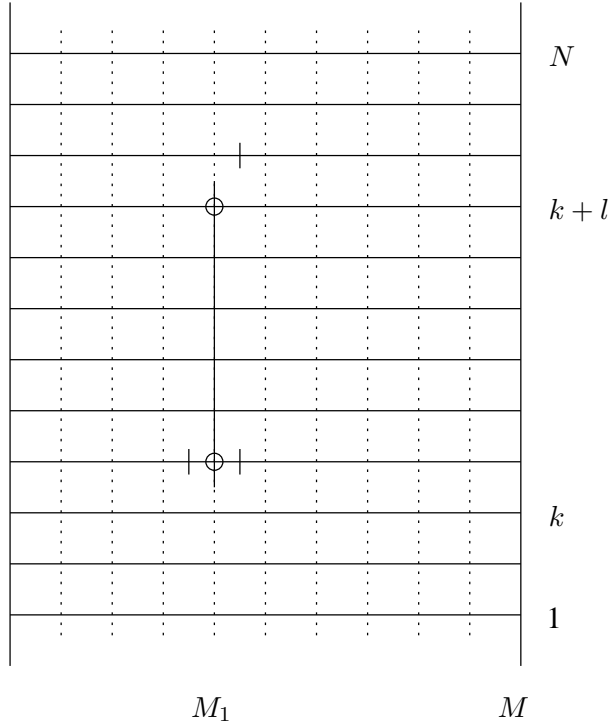,width=8cm}}
\caption{One loop correction to the propagator. The momenta
on the sites crossed by the solid vertical line are all
equal. The tick marks on the horizontal links just
after the interaction points indicate the deleted ghost
terms. The ghost terms associated with the circled sites
are coupled to the vertex insertions.}
\label{oneloopws}
\vskip.5cm
\end{figure}

So for the fission vertex, we choose to delete the corresponding
terms on both sides of the interaction point on the time
slice just after the fission, and this will
lead to a factor $1/l(M-l)$. By the same token if we delete 
a coupling term just after a fusion vertex and, say,
just to the right of the interaction point, we obtain the factor $1/M$. 
The fission vertex is represented by a time line on
which the momentum is constant. That is, the momentum on this
line is fixed (Dirichlet condition), but in a loop this fixed value is
integrated. Similarly, the ghost fields on this line are fixed
at zero.
As a concrete example, we give all the necessary factors
for the vertices, the ends of solid lines, in the context of the 
one loop correction to
the propagator, shown in Fig.~\ref{oneloopws}.

Since we require independent ghost variables
at the circled sites, we insist on the restriction $l\geq2$.
Applying the above considerations to this diagram, we have,
omitting for compactness the delta functions that implement
the Dirichlet conditions on the boundary $i=M$, 
\begin{eqnarray}
T^{\rm one~loop~II}_{fi}
&=&\int\prod_{j=1}^N\prod_{i=1}^{M-1}{dc_i^jdb_i^j\over2\pi}d^2q_i^j\ 
e^{-S}\ {g^2\over16\pi}\prod_{j=k+2}^{k+l}\delta({\bf q}_{M_1}^j
-{\bf q}_{M_1}^{j-1})\prod_{j=k+2}^{k+l-1}\left[2\pi\delta(b_{M_1}^j)\delta(c_{M_1}^j)\right]
\nonumber\\
&&\exp\left\{-{a\over m}\sum_{j=k+1}^{k+l}\left[2b_{M_1}^{j}c_{M_1}^{j}
-(b_{M_1-1}^{j}+b_{M_1+1}^{j})c_{M_1}^{j}-b_{M_1}^{j}
(c_{M_1-1}^{j}+c_{M_1+1}^{j})\right]\right\}\nonumber\\
&&\exp\left\{{a\over m}\left[b_{M_1}^{k+1}
c_{M_1}^{k+1}-b_{M_1-1}^{k+1}c_{M_1-1}^{k+1}-b_{M_1+1}^{k+1}c_{M_1+1}^{k+1}\right]\right\}\nonumber\\
&&\exp\left\{{a\over m}\left[b_{M_1}^{k+l}c_{M_1}^{k+l}
-(b_{M_1+1}^{k+l+1}-b_{M_1}^{k+l+1})
(c_{M_1+1}^{k+l+1}-c_{M_1}^{k+l+1})\right]\right\}
\label{oneloopi}
\end{eqnarray}
The superscript $II$ signifies that we have made minor changes
compared to Ref.~\cite{bardakcit}. In this new 
alternative, we associate factors
\bea
{\cal V}_0&\equiv& {ga\over4m\sqrt{\pi}}
\exp\left\{-{a\over m}(b_{M_1-1}^{k+1}c_{M_1-1}^{k+1}
+b_{M_1+1}^{k+1}c_{M_1+1}^{k+1})\right\}\\
{\bar{\cal V}}_0&\equiv& {ga\over4m\sqrt{\pi}}\exp\left\{-{a\over m}
(b_{M_1+1}^{k+l+1}-b_{M_1}^{k+l+1})
(c_{M_1+1}^{k+l+1}-c_{M_1}^{k+l+1})\right\}
\eea
with the fission and fusion vertices respectively, the different forms
due to the asymmetric way we have assigned $1/p^+$ factors to the
vertices. Note that since the factor on the second line of Eq.~\ref{oneloopi}
explicitly removes all the dependence on $b^j_{M_1}, c^j_{M_1}$
for $k+1\leq j\leq k+l$ from $S$, we are free to substitute
\bea
\prod_{j=k+2}^{k+l-1}\left[2\pi\delta(b_{M_1}^j)\delta(c_{M_1}^j)\right]
\to\exp\left\{2\pi\sum_{j=k+2}^{k+l-1} b^j_{M_1}c^j_{M_1}\right\},
\eea
which treats the interior solid line similarly to the way we
treated the boundary at $i=M$.
Then one can write the compact expression
\begin{eqnarray}
T^{\rm one~loop~II}_{fi}
&=&\int\prod_{j=1}^N\prod_{i=1}^{M-1}{dc_i^jdb_i^j\over2\pi}d^2q_i^j\ 
 \prod_{j=k+2}^{k+l}\delta({\bf q}_{M_1}^j
-{\bf q}_{M_1}^{j-1})\exp\left\{-S+b_{M_1}^{k+1}
c_{M_1}^{k+1}{\cal V}_0+b_{M_1}^{k+l}c_{M_1}^{k+l}{\bar{\cal V}}_0\right\}
\nonumber\\
&&\hskip-2.6cm\exp\left\{2\pi\sum_{j=k+2}^{k+l-1} b^j_{M_1}c^j_{M_1}
-{a\over m}\sum_{j=k+1}^{k+l}\left[2b_{M_1}^{j}c_{M_1}^{j}
-(b_{M_1-1}^{j}+b_{M_1+1}^{j})c_{M_1}^{j}-b_{M_1}^{j}
(c_{M_1-1}^{j}+c_{M_1+1}^{j})\right]\right\}
\label{oneloopii}
\end{eqnarray} 

The next step is to write a formula that systematically
sums over all the planar diagrams on the
lattice. The general planar diagram has
an arbitrary number of vertical solid lines. An interior link $j$ of
a solid line at $i=M_1$ is represented by the product of delta functions
\begin{eqnarray}
\delta({\bf q}_{i}^j
-{\bf q}_{i}^{j-1})
=\int {d{\bf y}_i^j\over(2\pi)^2} e^{i{\bf y}_i^j\cdot({\bf q}_{i}^j-
{\bf q}_{i}^{j-1})},
\end{eqnarray}
and by the $j$th term in the exponent 
of the last line of Eq.~\ref{oneloopii}.
A simple way to supply such factors is to assign an Ising spin $s_i^j
=\pm1$ to each site of the lattice. We assign $+1$ if the site $(i,j)$ 
is crossed by a vertical solid line, $-1$ otherwise. 
Then we can represent both kinds of link, solid line and dotted
line by the unified factor
\begin{eqnarray}
&&\int {d{\bf y}_i^j\over(2\pi)^2} 
\exp\left\{i{\bf y}_i^j
\cdot({\bf q}_{i}^j-{\bf q}_{i}^{j-1})P_i^jP_i^{j-1}
+\ln(4\pi^2/V)(1-P_i^jP_i^{j-1})\right\}\nonumber\\
&&\qquad\times\exp\left\{2\pi P_i^{j-1}P_i^jP_i^{j+1}b^j_{i}c^j_{i}
-{a\over m}P_i^j\left[2b_{i}^{j}c_{i}^{j}
-(b_{i-1}^{j}+b_{i+1}^{j})c_{i}^{j}-b_{i}^{j}
(c_{i-1}^{j}+c_{i+1}^{j})\right]\right\},
\end{eqnarray}
where we have defined the projector $P_i^j=(1+s_i^j)/2$
and where $V$ is the volume of transverse space.
Note that the way we have multiplied different terms by 
different projection operators ensures that the ranges
of summation are precisely as in our one-loop formula.

At the endpoints of each solid line we have to
supply the vertex insertions ${\cal V}_0$, ${\bar{\cal V}}_0$. 
These factors occur when
$s_i^j=-s_i^{j-1}$. So in the exponent we 
multiply the factors by $P_i^jP_i^{j+1}(1-P_i^{j-1})$,
$P_i^jP_i^{j-1}(1-P_i^{j+1})$ respectively. Again, the extra
projection operators are such as to suppress solid lines 
which cross only one horizontal line. That is, they enforce
the requirement that each internal loop 
occupies at least two time steps: $l\geq2$.

Our final formula for the sum of all planar diagrams is therefore
\begin{eqnarray}
T^{II}_{fi}&=&
\sum_{s_i^j=\pm1}\int DcDbD{\bf y}D{\bf q}
\exp\left\{{-S+\sum_{i,j}b_{i}^{j}
c_{i}^{j}P_i^j\left[{\cal V}_{0i}^jP_i^{j+1}(1-P_i^{j-1})
+{\bar{\cal V}}_{0i}^jP_i^{j-1}(1-P_i^{j+1})\phantom{\sum}\hskip-.5cm\right]}\right\}
\nonumber\\&&
\exp\left\{\sum_{i,j}\left[i{\bf y}_i^j\cdot({\bf q}_{i}^j-{\bf q}_{i}^{j-1})P_i^jP_i^{j-1}
+\ln(4\pi^2/V)(1-P_i^jP_i^{j-1})\right]\right\}\nonumber\\
&&\exp\left\{\sum_{i,j}P_i^j\left[2\pi P_i^{j-1}P_i^{j+1}b^j_{i}c^j_{i}
-{a\over m}\left[2b_{i}^{j}c_{i}^{j}
-(b_{i-1}^{j}+b_{i+1}^{j})c_{i}^{j}-b_{i}^{j}
(c_{i-1}^{j}+c_{i+1}^{j})\right]\right]\right\},
\label{isingsum}
\end{eqnarray}
where we have defined the measure as
\bea
DcDbD{\bf y}D{\bf q}\equiv\prod_{j=1}^N
\prod_{i=1}^{M-1}{dc_i^jdb_i^j\over2\pi}
{d{\bf y}_i^jd{\bf q}_i^j\over(2\pi)^2}.
\eea
The first exponent in this formula is just the action $S$ for the
free propagator plus the spin flipping
terms that take into account the creation and destruction of
solid lines. The second exponent takes care of the 
delta function insertions required for each solid line. The final
exponent adjusts the ghost couplings surrounding each
solid line. Notice that when $g=0$,
this exponential forces $T_{fi}$ to vanish
unless $s_i^j=s_i^{j-1}$ for all $i,j$,
because there would then not be enough ghost factors to
saturate the ghost integrals. Thus solid lines are
eternal if $g=0$, corresponding to the free field case.

We remark that the expression (\ref{isingsum}) sums all the
planar multi-loop corrections to the propagator of the
matrix scalar field. The evolving system is therefore in the
adjoint representation of the color group. That is, we have tacitly
assumed that the only solid lines initially and finally are
those at the boundaries of the strip. More general initial and
final states are described by allowing more solid lines initially
and finally. When the system is in a color singlet state,
we must of course include diagrams in which the outer boundaries
are identified, \ie\ the diagrams should be drawn on a
cylinder, not a strip. In this case, 
${\bf q}(p^+)={\bf q}(0)+{\bf p}$, the variable ${\bf q}(\sigma)$
is strictly periodic only in the state of zero
total transverse momentum. 

Finally, we mention an approximate treatment of the momentum
delta functions on the solid lines that may be more suited
to practical calculations. The idea is to use a Gaussian
approximation for the delta function:
\bea
\delta({\bf q})=\lim_{\epsilon\to0}{1\over(2\pi\epsilon)^{d/2}
}\exp\left\{-{{\bf q}^2\over2\epsilon}\right\}.
\eea
We can keep $\epsilon$ finite until the end of the calculation.
However, if we want to recover the exact result
of using delta functions, we should send $\epsilon\to0$ {\it before} the
continuum limit $a,m\to0$. Using this device, our formula for
the sum of planar diagrams becomes (for $d=2$):
\begin{eqnarray}
T^{II}_{fi}&=&\lim_{\epsilon\to0}
\sum_{s_i^j=\pm1}\int DcDbD{\bf q}
\exp\left\{{-S+\sum_{i,j}b_{i}^{j}
c_{i}^{j}P_i^j\left[{\cal V}_{0i}^jP_i^{j+1}(1-P_i^{j-1})
+{\bar{\cal V}}_{0i}^jP_i^{j-1}(1-P_i^{j+1})\right]}\right\}
\nonumber\\&&
\exp\left\{-\sum_{i,j}\left[{({\bf q}_{i}^j-{\bf q}_{i}^{j-1})^2\over2\epsilon}
+\ln(2\pi\epsilon)\right]P_i^jP_i^{j-1}\right\}\label{isingsumepsilon}\\
&&\exp\left\{\sum_{i,j}P_i^j\left[2\pi P_i^{j-1}P_i^{j+1}b^j_{i}c^j_{i}
-{a\over m}\left(2b_{i}^{j}c_{i}^{j}
-(b_{i-1}^{j}+b_{i+1}^{j})c_{i}^{j}-b_{i}^{j}
(c_{i-1}^{j}+c_{i+1}^{j})\right)\right]\right\},\nonumber
\end{eqnarray}
with the measure now given by
\bea
DcDbD{\bf q}\equiv\prod_{j=1}^N
\prod_{i=1}^{M-1}{dc_i^jdb_i^j\over2\pi}
{d{\bf q}_i^j}.
\eea
\section{Yang-Mills}
\subsection{Treatment of Factors of Transverse Momentum}
Here, we use the notation and conventions in Ref.~\cite{beringrt},
according to which the values of the 
non-vanishing three transverse gluon vertices are:
\bea
{{}\atop\mbox{\epsfig{
file=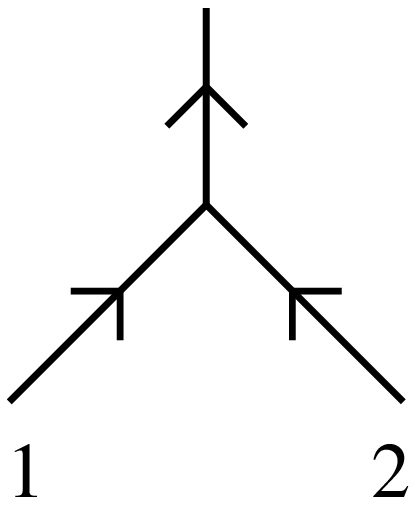,width=1.5cm}}}
\displaystyle\quad&=&{{ga\over 4m\pi^{3/2}}M_3
\left({p_2^\wedge\over M_2}-{p_1^\wedge\over M_1}\right)}
\label{upupdown}\\
{{}\atop\mbox{\epsfig{
file=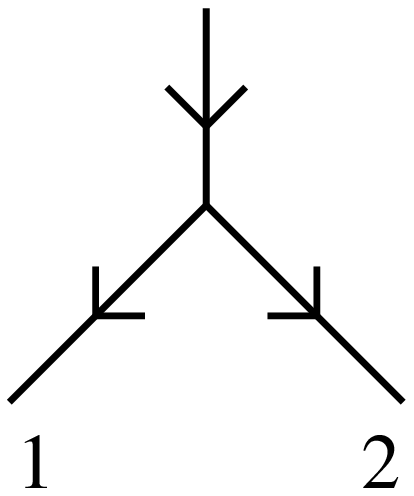,width=1.5cm}}}
\displaystyle\quad&=&{{ga\over 4m\pi^{3/2}}M_3
\left({p_2^\vee\over M_2}-{p_1^\vee\over M_1}\right)}
\label{downdownup}
\eea 
Here, $p^\wedge=(p^x+ip^y)/\sqrt2$, $p^\vee=(p^x-ip^y)/\sqrt2$, 
and $M_im$ is the discretized
$p^+$ {\it entering} the diagram on leg $i$. These are the traditional
Feynman diagram factors multiplied by $a/8m\pi^{3/2}$ which comes
from our rearrangement of propagator and vertex assignments.
The $1/p^+$ factors $1/|M_3|$ or $1/|M_1M_2|$ are not included. 
We remind the reader that these are light-cone gauge ($A_-=0$) expressions
and include the contributions that arise when the longitudinal
field $A_+$ is eliminated from the formalism.

The basic expressions for the complete quartic vertices,
combining those from ``Coulomb'' exchange with those
from the $\Tr[A_i,A_j]^2$ term in the Lagrangian, can be
manipulated into a form that suggests a concatenation of
two cubics. In Cartesian basis we can write them in the following way:
\bea
\Gamma^{i_1i_2i_3i_4}&=&{g^2 a\over 32m\pi^3}\left\{\delta_{i_1i_2}\delta_{i_3i_4}
{(M_1-M_2)(M_4-M_3)\over(M_1+M_2)^2}+\left(\delta_{i_1i_3}\delta_{i_2i_4}
-\delta_{i_1i_4}\delta_{i_2i_3}\right)\right\}\nonumber\\
&&+{g^2 a\over 32m\pi^3}\left\{\delta_{i_2i_3}\delta_{i_1i_4}
{(M_1-M_4)(M_2-M_3)\over(M_1+M_4)^2}+\left(\delta_{i_1i_3}\delta_{i_2i_4}
-\delta_{i_1i_2}\delta_{i_3i_4}\right)\right\}
\eea
Where as before all $M_k$ are taken to flow {\it into} the vertex,
and the rearrangement factor $a/32m\pi^3$ has also been included.
The first line on the r.h.s. looks like the exchange of an
$O(d)$ singlet and an $O(d)$ antisymmetric tensor in the $s$
channel (12,34) and the second line like the same exchanges in the
$t$ channel (23,41). It is easy to confirm in the case $d=2$ that 
these expressions are summarized in complex basis by the
pair of diagrams 
\bea
{{}\atop\mbox{\epsfig{
file=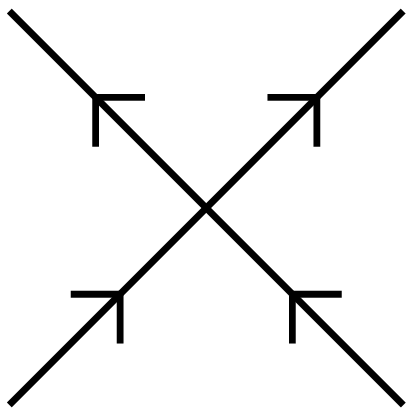,width=1.2cm}}}
\displaystyle\quad&=&{{g^2a\over 32m\pi^3}\left({(M_1-M_4)(M_2-M_3)\over
(M_1+M_4)^2}+1\right)
\label{upupdowndown}}\\
{{}\atop\mbox{\epsfig{
file=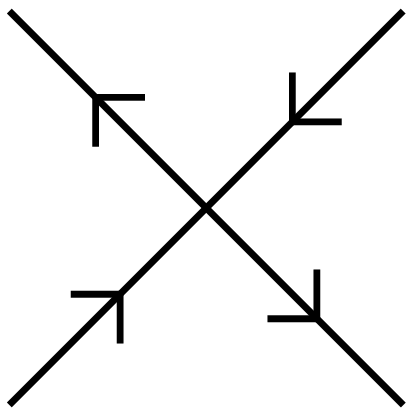,width=1.2cm}}}
\displaystyle\quad&=&{{g^2a\over 32m\pi^3}}\left({(M_1-M_4)(M_2-M_3)\over
(M_1+M_4)^2}+{(M_1-M_2)(M_4-M_3)\over
(M_1+M_2)^2}-2\right)
\label{updownupdown}
\eea 

\begin{figure}[ht]
\begin{center}
${\epsfig{file=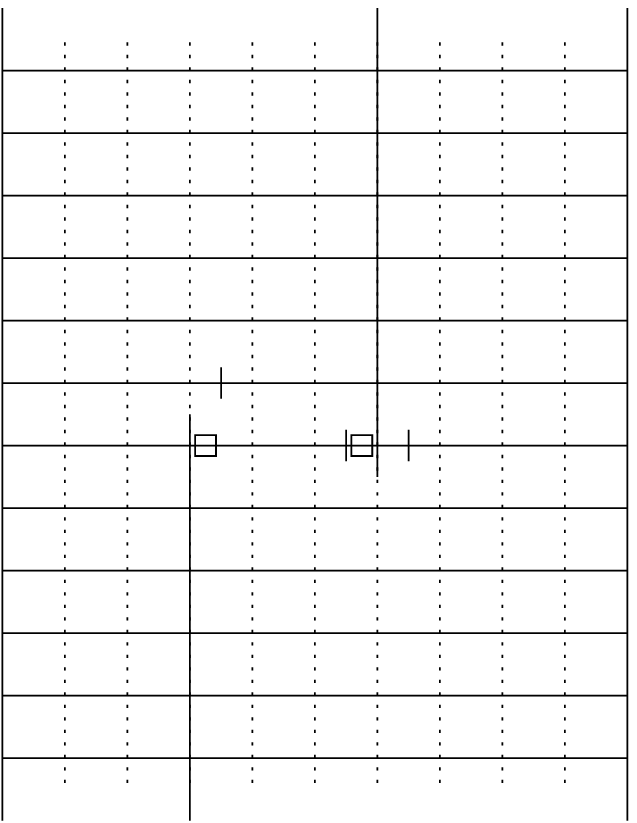,width=4cm}\atop {\displaystyle\rm a)}}
\hskip1cm\quad
{\epsfig{file=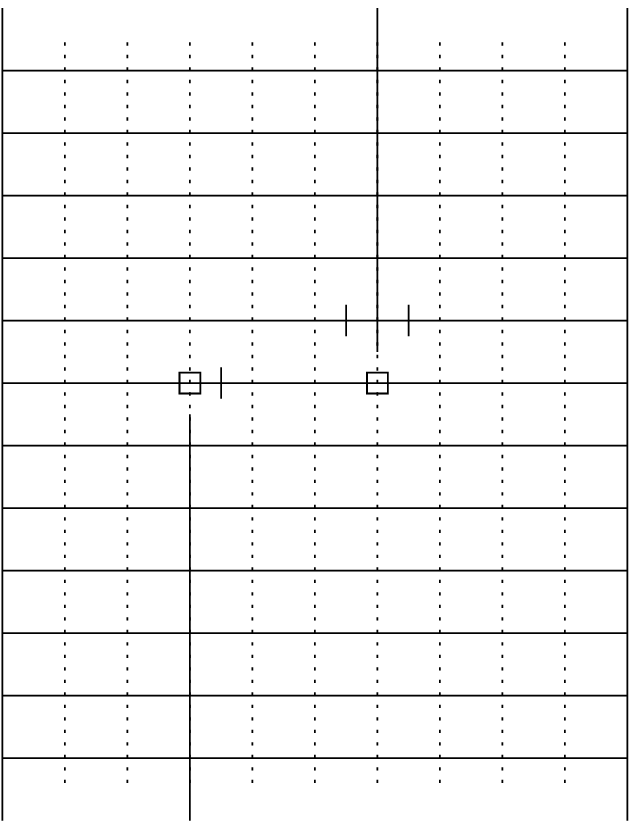,width=4cm}\atop {\displaystyle\rm b)}}
\hskip1cm\quad
{\epsfig{file=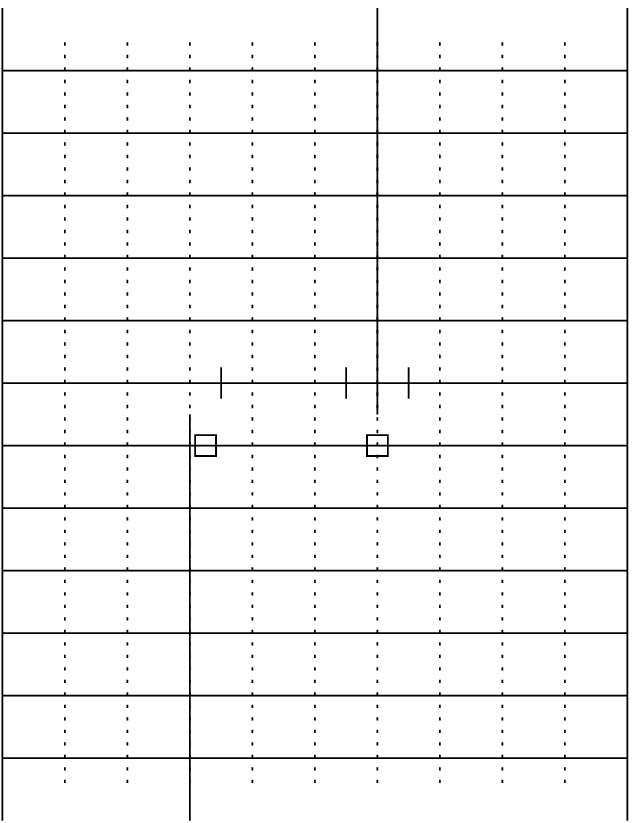,width=4cm}\atop {\displaystyle\rm c)}}
$
\end{center}
\caption{Possible contributions to the quantum term from two
$\Delta{\bf q}$ insertions placed at the location of the
open squares on the same time slice. Figures a) and b) produce 
the desired quartic vertices. However c) is an artifact
of our insertion procedure and must be suppressed. The tick marks
identify the deleted ghost terms according to light-cone priority.
The double deletion on strip 4 in figure c) provides a zero that
suppresses this artifact.}
\label{contact231}
\end{figure}

We next explain how the transverse momentum
factors can be handled in our worldsheet construction. 
Recall that our scheme for
spreading the propagator among $M$ bits, involved the integral
\begin{eqnarray}
I=\int d^2q_1\cdots d^2q_{M-1}\exp\left\{-{a\over2m}\sum_{i=0}^{M-1}
({\bf q}_{i+1}-{\bf q}_{i})^2\right\}.
\end{eqnarray}
It is not hard to show that insertion in the integrand of the
factor $[{\bf q}_l-{\bf q}_{l-1}]$, where $l$ labels the $l$th
bit leads to:
\begin{eqnarray}
\int d^2q_1\cdots d^2q_{M-1}\ [{\bf q}_l-{\bf q}_{l-1}]\ 
\exp\left\{-{a\over2m}\sum_{i=0}^{M-1}
({\bf q}_{i+1}-{\bf q}_{i})^2\right\}={{\bf q}_M-{\bf q}_0\over M}I.
\end{eqnarray}
This is an identity for any $l=1,2,\ldots M$.
For later use we quote the generating functions
\bea
\left\langle \exp\left\{\sum_{i=1}^{M-1} J_iq_i\right\}\right\rangle
&=&\exp\left\{{m\over 2a}\sum_i {i(M-i)\over M}J_i^2
+{m\over a}\sum_{i<j}{i(M-j)\over M}J_iJ_j\right.\nonumber\\
&&\hskip3cm\left.+{q_M\over M}\sum_i iJ_i
+{q_0\over M}\sum_i(M-i)J_i\right\}\\
\left\langle \exp\left\{\sum_{i=0}^{M-1} 
K_i(q_{i+1}-q_i)\right\}\right\rangle
&=&\exp\left\{{m\over 2a}\sum_i K_i^2
-{m\over 2aM}\sum_{i,j}K_iK_j+{q_M-q_0\over M}\sum_i K_i
\right\}
\eea
From the second of these we see immediately 
that two vertex insertions on the
same time slice and on the same strip
produce a ``quantum fluctuation'' contribution:
\bea
\left\langle (q_{i+1}-q_i)(q_{j+1}-q_j)\right\rangle
=\left({q_M-q_0\over M}\right)^2
+{m\over a}\left[\delta_{ij}-{1\over M}\right]\; .
\eea
These fluctuation terms cause two coincident cubics to behave
as a quartic contact vertex, and must be carefully analyzed
in the context of translating the Yang-Mills quartic
vertices into our worldsheet language. Because the
$\Delta{\bf q}$ insertions on the three different
worldsheet strips joined at a vertex 
are applied on two different time slices, we
have the three possible contributions shown in Fig.~\ref{contact231}.
The a) and b) contributions lead to the 
quartic vertices required by the Yang-Mills Feynman
rules.

\begin{figure}[ht]
\begin{center}
$\epsfig{file=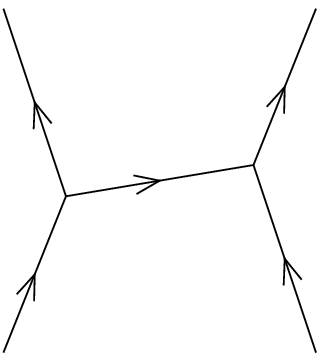,width=2cm}\hskip2cm
\epsfig{file=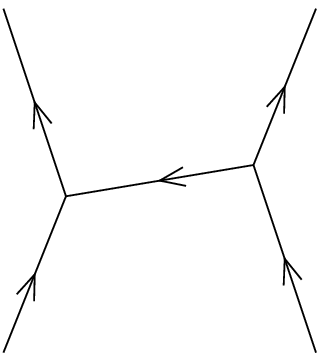,width=2cm}$
\caption{A four gluon amplitude}
\label{fourgluons}
\end{center}
\end{figure}

To see how this works in a simple example, 
consider the four gluon diagrams built from two cubics
in Fig.~\ref{fourgluons}. 
The product of vertex factors is
\bea
{g^2a^2\over 16m^2\pi^3}M_2M_4\left(
{p_2^\vee+p_3^\vee\over M_2+M_3}-{p_3^\vee\over M_3}\right)
\left({p_1^\wedge+p_4^\wedge\over M_1+M_4}-{p_1^\wedge\over M_1}\right).
\eea
In this process a fluctuation contribution of the
type shown in Fig.~\ref{contact231}a) comes from
a double insertion on the intermediate string
with momentum $p_1+p_4=-p_2-p_3$. Remembering the $1/|M_1+M_4|$
factor from the intermediate propagator,
the contribution of the quantum term is
\bea
{\rm Quantum~Term} = -2{g^2a\over 32m\pi^3}{M_2M_4+M_1M_3\over(M_1+M_4)^2},
\label{quantumterm}
\eea
where the $M_1M_3$ term comes from the diagram where the
arrow on the intermediate line is reversed. Then one observes
\bea
-2(M_2M_4+M_1M_3)=(M_1-M_4)(M_2-M_3)+(M_1+M_4)^2.
\eea
The first term inserted in the numerator of Eq.~\ref{quantumterm}
yields precisely the quartic
vertex contribution from the induced instantaneous ``Coulomb''
exchange, and the second term yields precisely
the contribution of the commutator squared term in the 
Yang-Mills Lagrangian (see Eq.~\ref{upupdowndown}). 
It is not hard to confirm that the
a) and b) type contributions correctly produce the
quartic vertices for all other spin assignments to the external gluons. 

On the other hand contact interactions arising from contributions
like Fig.~\ref{contact231}c) are spurious and should not be
included. We see that these spurious contributions arise
when there is zero propagation time for the
gluon exchanged between the two cubic vertices. The need to
remove these spurious contributions is a strong argument to
use our original ghost insertion scheme to account
for $1/p^+$ factors. Then the troublesome
contact interaction is accompanied  by the
deletion of {\it two} ghost terms on the same time slice, which
supplies a welcome zero. This doesn't happen for an
alternative ghost insertion scheme based on spin flow mentioned
later.

Now let's consider the worldsheet representation of the vertices
of Eqs.~\ref{upupdown},~\ref{downdownup} in more detail. In the fusion case 
where 1 and 2 are incoming and 3 is outgoing.
Let gluon 1 have $M_1=l$ and gluon 2 have $M_2=M-l$. The vertex
is represented by a strip of width $M$, the left boundary
at $i=0$ and the right boundary at $i=M$ and  with a solid line coming
in at $i=l$. Consider the time slice $j=k$ just before the interaction
point. Represent the insertion of a factor $q^j_i$ by the notation
$\langle q^j_i\rangle$. Then we have
\begin{eqnarray}
\langle q^k_{l+1}-q^k_l\rangle={q_M-q_l\over M-l}={p_2\over M_2}\\
\langle q^k_l-q^k_{l-1}\rangle={q_l-q_0\over l}={p_1\over M_1}
\end{eqnarray}
Thus insertion of the factor $q_{l+1}+q_{l-1}-2q_l$ yields
the desired vertex factor 
\begin{eqnarray}
\langle (q^k_{l+1}+q^k_{l-1}-2q^k_l)^\wedge\rangle
=\left({p_2^\wedge\over M_2}-{p_1^\wedge\over M_1}\right).
\end{eqnarray}
When 3 and 1 are incoming, we have to put one insertion on the
time slice $k+1$ and the other at the time slice $k$: 
\begin{eqnarray}
\langle q^k_{l+1}-q^k_l\rangle={q_M-q_l\over M-l}={p_1\over M_1}\\
\langle q^{k+1}_{l+1}-q^{k+1}_l\rangle={q_M-q_0\over M}={p_2\over M_2}\\
\langle q^{k+1}_{l+1}-q^{k+1}_l-q^k_{l+1}+q^k_l\rangle={p_2\over M_2}
-{p_1\over M_1}.
\end{eqnarray}
Similarly, when 2 and 3 are incoming
\bea
\langle q^{k}_{l}-q^{k}_{l-1}-q^{k+1}_{l}+q^{k+1}_{l-1}\rangle={p_2\over M_2}
-{p_1\over M_1}.
\eea
The fission cases where with one particle incoming and two outgoing are
treated analogously.

\subsection{Treatment of Remaining $p^+$ Factors}
In order to suppress the spurious contact
contributions of Fig.~\ref{contact231}c), we choose the
same ghost modifications at the vertex (to handle
$1/p^+$ factors) as in the
scalar case (see Fig.~\ref{oneloopws}).
Then the factor of $M_3$ of Eqs~(\ref{upupdown},~\ref{downdownup})
still needs to be represented locally on the world sheet.
This seems to be very difficult unless we introduce
yet more ghost degrees of freedom. One way to do this
is to note that we are always free to introduce two
``dummy'' ghost systems, $\beta,\gamma$ and ${\bar\beta},{\bar\gamma}$ 
whose net effect is
to include a factor of unity in the path integral.
On each time slice for each strip we choose the
action as in Eq.~\ref{ghostunity1} for $\beta,\gamma$ and the one 
in Eq.~\ref{ghostunity2} for ${\bar\beta},{\bar\gamma}$. 
In other words on each time slice we insert
\begin{eqnarray}
&&\int \prod_{i=1}^{M-1} {d\gamma_id\beta_i}
{d{\bar\gamma}_id{\bar\beta}_i} 
\exp\left\{
\beta_1\gamma_1+\sum_{i=1}^{M-2}
(\beta_{i+1}-\beta_i)(\gamma_{i+1}-\gamma_i)\right.\nonumber\\
&&\hskip4cm\left.+
{\bar\beta}_{M-1}{\bar\gamma}_{M-1}
+\sum_{i=1}^{M-2}({\bar\beta}_{i+1}-{\bar\beta}_i)
({\bar\gamma}_{i+1}-{\bar\gamma}_i)\right\}
=1.
\label{dummyunity}
\end{eqnarray}
Here we have dispensed with the factors of $2\pi$ and $a/m$
present in the $b,c$ path integral.
This insertion is completely harmless because it does nothing. 
But with these dummy-ghost systems 
available, we can locally produce factors of $M_i$
at will as they are needed. 
For example, either $e^{\beta_{M-1}\gamma_{M-1}}$ applied
on the right of a strip of $M$ bits 
or $e^{{\bar\beta}_1{\bar\gamma}_1}$ applied
on the left of the strip produces a factor of $M$. 
At a vertex, the end of
a solid line marks where two strips, 1 to the left of 2, 
join a single larger strip
3. Then an insertion of the first type on strip 1 produces $M_1$, of the
second type on strip 2 produces $M_2$, and the sum of the two  
insertions produces $M_1+M_2=-M_3$. This scheme is perhaps
over-generous in the number of dummy-ghosts introduced,
but it serves to prove that a local description is
possible.

As an aside, we briefly mention an alternative treatment of $p^+$ factors
assigned to vertices on the basis of spin rather than
light-cone priority. In the
complex basis $(A_1\pm iA_2)/\sqrt2$ each transverse
propagator carries an arrow denoting the flow of
spin. Suppose we associate the $1/2|p^+|$ factor with
the vertex at, say, the {\it tail}. Then there will always be 
enough denominator factors at each vertex to cancel
all $p^+$ factors in the numerator, leaving {\it at most}
one $p^+$ factor in the denominator.
Again take $p_k=M_km$ to be the momentum
flowing {\it in} to the vertex, so $p^+_k>0(<0)$ if $k$ is
incoming (outgoing). The vertex in Eq.~\ref{upupdown}
will then receive the factor $1/|M_3|$, so it will be assigned
the value
\bea
\mbox{\epsfig{file=3VertexUpUpDown.eps,width=1.5cm}}^{{}_{
\displaystyle\quad{\to{\rm sgn}(p^+_3){ga\over 4m{\pi^{3/2}}}
\left({p_2^\wedge\over M_2}-{p_1^\wedge\over M_1}\right)}}}
\eea 
and will have {\it no} leftover $p^+$ factors.
Contrast this with the vertex in Eq.~\ref{downdownup},
which receives the factor $1/|M_1M_2|$:
\bea
\mbox{\epsfig{file=3VertexDownDownUp.eps,width=1.5cm}}^{{}_{
\displaystyle\quad{\to-{\rm sgn}(p^+_1p^+_2)
{ga\over 4m\pi^{3/2}}\left({1\over M_1}+{1\over M_2}\right)
\left({p_2^\vee\over M_2}-{p_1^\vee\over M_1}\right).}}}
\eea
Here the $p^+$ factors are always in the 
denominator and can thus be produced by modifying the $b, c$ 
ghost action near the interaction point. The virtue of
this alternative scheme is that all factors of $p^+$
are accounted for without introducing extra ghosts. However,
it leaves the problem of eliminating the spurious
contact quartics arising from Fig.~\ref{contact231}c) in
some other way,
because this alternate ghost arrangement provides no zero for that
diagram. Diagram by diagram, one could discard it by
hand, but to do it dynamically probably requires
the introduction of extra ghost variables.

\subsection{Polarization Flow Described by a Worldsheet Grassmann Field}
To keep track of the flow of polarization indices for large diagrams
we would like to introduce some worldsheet fermions, 
which in the absence of interactions
do nothing more than reproduce the Kronecker delta, but
in multi-loop diagrams reproduce the sum over all possible
polarizations of internal lines.

We introduce worldsheet spinors of the
Neveu-Schwarz-Ramond type, ${\bf S}_k$.
Let us first evaluate some simple Grassmann integrals. Consider
a single chain ${\bf S}_1,{\bf S}_2,\ldots, {\bf S}_{2K-1},{\bf S}_{2K}$,
with an  {\it even} number of spin variables, each carrying a transverse
vector index and use the action
\bea
A=\sum_{i=1}^{2K-1}{\bf S}_i\cdot{\bf S}_{i+1}.
\eea
We shall define the measure for integration of the
Grassmann variables by
\bea
{\cal D}S\equiv\prod_{i=1}^d\left[dS_{2K}^idS_{2K-1}^i\cdots dS_1^i\right]
\eea
Then it is easy to check that
\bea
\int {\cal D}S\ e^A&=&1\\
\int{\cal D}S\  e^{A+{\bf\eta}_1\cdot{\bf S}_1
+{\bf\eta}_{2K}\cdot{\bf S}_{2K}}
&=&e^{{\bf\eta}_{2K}\cdot{\bf\eta}_1}.
\eea
In particular, the last equation implies
\bea
\int{\cal D}S\ S_1^k\ S_{2K}^l\ e^A&=&\delta_{kl}.
\label{kdeltao}
\eea
Note that these formulae {\it require} an even number of spins.

It is not hard to show that with an odd number of spins ($2K+1$ say)
\bea
\int {\cal D}S\ e^A&=&0\qquad{\rm for}~2K+1~{\rm spins}\\
\int{\cal D}S\  e^{A+{\bf\eta}_1\cdot{\bf S}_1
+{\bf\eta}_{2K+1}\cdot{\bf S}_{2K+1}}
&=&\pm\prod_{i=1}^d(\eta_1^i+\eta_{2K+1}^i),
\eea
with the sign depending on the specific measure convention.
In our worldsheet construction, we shall find that in
order to guarantee the consistent application of the
even spin formulae in the presence of interactions, the
number of spins assigned to each bit must be a multiple of four.

\begin{figure}[ht]
\centerline{\epsfig{file=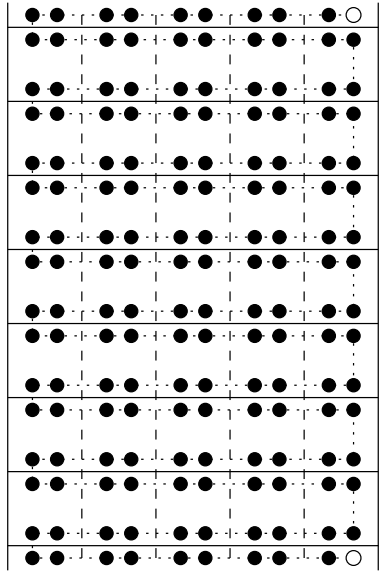,width=5cm}}
\caption{Assignment of Grassmann spins for propagator. Each dot is assigned
a Grassmann spin ${\bf S}$ and the bond pattern for the spin chain is
indicated by dotted lines.}
\label{gluonprop}
\end{figure}

For the propagator, we can assign one of the spin variables to each
of the dots open or closed in Fig.~\ref{gluonprop}. The open circles are
where we will insert a factor $S^{i}$ or $S^j$, so that the
Grassmann integrals will yield the required Kronecker delta. It is
interesting to write the formal continuum limit of the bulk part
of the action to use in the propagator path integral. To do this
we rename the four spins living on the $i$th bit on time slice
$j$ clockwise starting on the
lower right hand corner: ${\bf S}^j_{2 i}$,  ${\bar{\bf S}}^j_{2 i}$, 
${\bf S}^j_{1 i}$, ${\bar{\bf S}}^j_{1 i}$, respectively.
Then, using the bond pattern implied by our spin
chains, we find the bulk spin  action
\bea
S_s=-\sum_{ij} {\bar{\bf S}}^j_{1 i}\cdot({\bf S}^j_{1 i+1}
-{\bf S}^j_{1 i})
+\sum_{ij} {\bar{\bf S}}^j_{2 i}\cdot({\bf S}^j_{2 i}
-{\bf S}^j_{2 i-1}).
\label{spinaction}
\eea 
In this form we easily read off the formal continuum limit.
First identify continuum spin variables by the
rule ${\bf S}^j_{(1,2)i}\to\sqrt{a}{\bf S}_{1,2}(\sigma,\tau)$,
and similarly for ${\bar{\bf S}}_{1,2}$. Then we
have the formal continuum limit
\bea
S_s\to \int d\tau d\sigma({\bar{\bf S}}_2\cdot{\bf S}_2^\prime
-{\bar{\bf S}}_1\cdot{\bf S}_1^\prime).
\eea
where the prime denotes $d/d\sigma$. Note that although the
bulk has a simple continuum limit, the accurate dynamical
treatment at the boundaries heavily relies on the $x^+,p^+$ lattice. 

\begin{figure}[ht]
\psfrag{'k'}{$k$}
\psfrag{'l'}{$l$}
\centerline{\epsfig{file=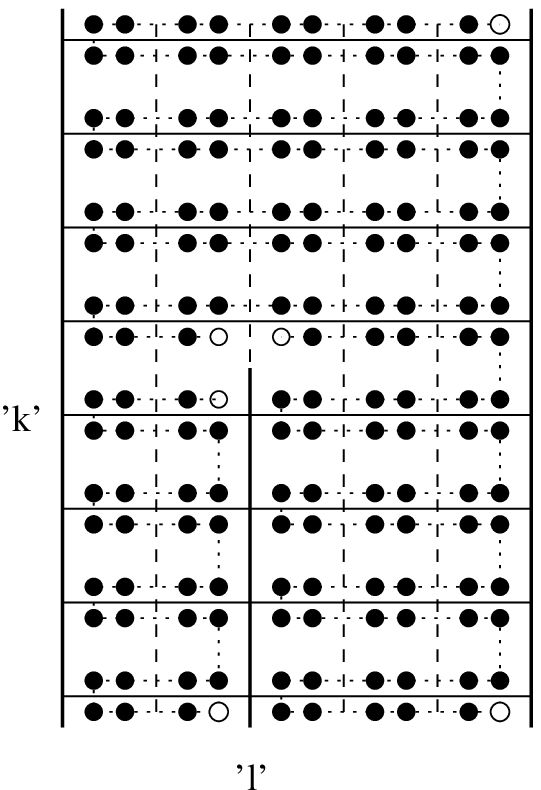,width=5cm}
\hskip1in\epsfig{file=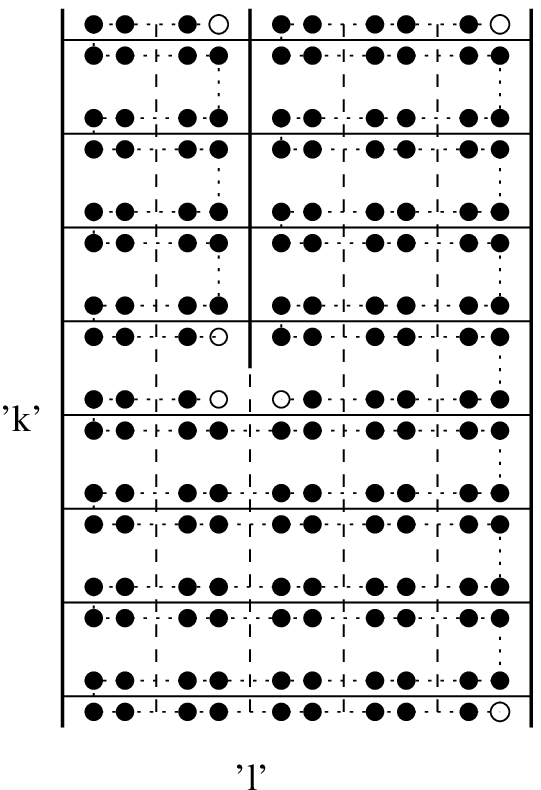,width=5cm}
}
\caption{Cubic fusion and fission vertices.}
\label{cubicvert}
\end{figure} 

Next we turn to interactions. 
We draw the diagrams for the fusion and fission 
three gluon vertices in Fig.~\ref{cubicvert}.
First consider the fusion vertex. 
Label the external legs  
1, 2, 3, ordered counter-clockwise starting at the lower left. 
The three open circles mark the spins that participate in the
vertex insertion that should yield the correct Yang-Mills 
cubic vertex. Note that the insertion will be
cubic in spin variables and hence Grassmann odd. This is
necessary because we have inserted a single spin variable
on each external line to represent the
polarization of the external gluon.
In the fusion case, starting at the lowest point and working
counter-clockwise around the triangle, call these spins
${\bf S}_A$, ${\bf S}_B$, and ${\bf S}_C$ respectively. Finally,
let $k$ label the time slice just before the solid
line ends and let $l$ label the spatial location of the
solid line. Then, with ghost modifications according to light-cone 
priority, the vertex insertion will be
\bea
{{\bar{\cal V}}\over2\pi}=\left({\bf u}\cdot{\bf S}_A{\bf S}_B\cdot{\bf S}_C
+{\bf v}\cdot{\bf S}_B{\bf S}_C\cdot{\bf S}_A
+{\bf w}\cdot{\bf S}_C{\bf S}_A\cdot{\bf S}_B\right)\exp\left\{
-{a\over m}(b_{l+1}^{k+1}-b_{l}^{k+1})(c_{l+1}^{k+1}-c_{l}^{k+1})\right\}.
\label{yangmillsbarv}
\eea
Here ${\bf u}$, ${\bf v}$, and ${\bf w}$ can be determined
from Eqs.~\ref{upupdown}, \ref{downdownup} by writing out the
coefficient on the r.h.s. in complex basis\footnote{Incidentally, 
referring to Eq.~\ref{veecomplex}, 
the rule using the spin to dictate the assignment of 
the $1/p^+$ factors from the propagators can be 
simply stated: $S_k^\wedge\to S_k^\wedge/|M_k|$
but $S_k^\vee\to S_k^\vee$.}:
\bea
&&\hskip-2cm{\bf u}\cdot{\bf S}_A{\bf S}_B\cdot{\bf S}_C
+{\bf v}\cdot{\bf S}_B{\bf S}_C\cdot{\bf S}_A
+{\bf w}\cdot{\bf S}_C{\bf S}_A\cdot{\bf S}_B\nonumber\\
&&={(u+v)^\wedge}{S^\vee}_A{S^\vee}_B{S^\wedge}_C
+{(u+v)^\vee}{S^\wedge}_A{S^\wedge}_B{S^\vee}_C
+{(v+w)^\wedge}{S^\wedge}_A{S^\vee}_B{S^\vee}_C\nonumber\\
&&+{(v+w)^\vee}{S^\vee}_A{S^\wedge}_B{S^\wedge}_C
+{(u+w)^\wedge}{S^\vee}_A{S^\wedge}_B{S^\vee}_C
+{(u+w)^\vee}{S^\vee}_A{S^\wedge}_B{S^\vee}_C
\label{veecomplex}
\eea
from which we infer 
\bea
{\bf u}+{\bf v}&=&{ga\over4m{\pi^{3/2}}}p^+_3\left({{\bf p}_2\over p^+_2}-
{{\bf p}_1\over p^+_1}\right)\\
{\bf v}+{\bf w}&=&{ga\over4m{\pi^{3/2}}}p^+_1\left({{\bf p}_3\over p^+_3}-
{{\bf p}_2\over p^+_2}\right)\\
{\bf w}+{\bf u}&=&{ga\over4m{\pi^{3/2}}}p^+_2\left({{\bf p}_1\over p^+_1}-
{{\bf p}_3\over p^+_3}\right).
\eea
As already stated, the ${{\bf p}_i/p_i^+}$ can be
produced by insertions of $\Delta{\bf q}$ at strategic locations
near the interaction point:
\bea
{{\bf p}_1\over M_1}\to{\bf q}^k_l-{\bf q}^k_{l-1}\ ,\qquad
{{\bf p}_2\over M_2}\to{\bf q}^k_{l+1}-{\bf q}^k_{l}\ ,\qquad
{{\bf p_3}\over M_3}\to{\bf q}^{k+1}_l-{\bf q}^{k+1}_{l-1}.
\eea
The remaining factors of $p^+_i$ can then be supplied by 
dummy ghost insertions 
\bea
M_1\to e^{\beta^k_{l-1}\gamma^k_{l-1}}, \qquad
M_2\to e^{{\bar\beta}^k_{l+1}{\bar\gamma}^k_{l+1}}, \qquad
M_3\to-e^{\beta^k_{l-1}\gamma^k_{l-1}}
- e^{{\bar\beta}^k_{l+1}{\bar\gamma}^k_{l+1}}
\eea

For the fission case, starting from the bottom left open circle of
the triangle and working around the triangle counter-clockwise,
call the inserted spins
${\bf S}_D$, ${\bf S}_E$, and ${\bf S}_F$ respectively. Also the
{\it incoming} momenta are assigned counter-clockwise, starting
with the bottom leg, $p_1$, $p_2$, $p_3$ respectively.
Then the vertex insertion will have the similar structure
\bea
{{\cal V}\over2\pi}=\left({\bf u}\cdot{\bf S}_D{\bf S}_E\cdot{\bf S}_F
+{\bf v}\cdot{\bf S}_E{\bf S}_F\cdot{\bf S}_D
+{\bf w}\cdot{\bf S}_F{\bf S}_D\cdot{\bf S}_E\right)
\exp\left\{
-{a\over m}(b_{l+1}^{k+1}c_{l+1}^{k+1}+b_{l-1}^{k+1}c_{l-1}^{k+1})\right\}
\label{yangmillsv}
\eea
with the same expressions for ${\bf u}$, ${\bf v}$, and ${\bf w}$,
but a different representation of the $p^+_i$ factors
\bea
M_3\to -e^{\beta^{k+1}_{l-1}\gamma^{k+1}_{l-1}}, \qquad
M_2\to -e^{{\bar\beta}^{k+1}_{l+1}{\bar\gamma}^{k+1}_{l+1}}, \qquad
M_1\to e^{\beta^{k+1}_{l-1}\gamma^{k+1}_{l-1}}
+e^{{\bar\beta}^{k+1}_{l+1}{\bar\gamma}^{k+1}_{l+1}}
\eea

For the purposes of writing a formula for the sum of all planar diagrams,
as in Eq.~\ref{isingsum}, it is better to use a Grassmann
even version of the vertex insertion. A simple way to do this is
to add a component $S^{d+1}$ to the spin variables ${\bf S}$,
making the spin a vector in $d+1$ dimensions. Then $S^{d+1}$ can
be used to represent the Kronecker delta with even variables
(see Eq.~\ref{kdeltao}):
\bea
\int DS S_1^kS_1^{d+1}S_{2K}^{d+1}S_{2K}^l=\delta_{kl}.
\eea 
When one uses the even insertions $S^kS^{d+1}$ on incoming
external lines and $S^{d+1}S^{k}$ on outgoing external lines,
then everything goes through as before provided
we use the Grassmann even vertex insertions.
\bea
{\cal V}_e\equiv {\cal V}S_D^{d+1}S_E^{d+1}S_F^{d+1}\ ,\qquad
{\bar{\cal V}}_e\equiv S_A^{d+1}S_B^{d+1}S_C^{d+1}{\bar{\cal V}}.
\label{evenvees}
\eea
Note that we have normalized ${\cal V}_e$ and $\bar{\cal V}_e$ so that
they will appear in the expression for the sum of all planar diagrams
with exactly the same coefficients as with the scalar case. 
\section{Concluding Remarks}
In this article we have shown how to extend the worldsheet
construction of Ref.~\cite{bardakcit} to the case of Yang-Mills
theories. This extension requires the introduction of
a worldsheet fermionic spin variable ${\bf S}$ to
account for the flow of gluon spin through complicated
diagrams.

Although we have not explicitly written out the 
rather unwieldy analog for Yang-Mills of Eq.~\ref{isingsum},
the formula summing planar diagrams, 
we have completely specified all of
its ingredients.
To construct the corresponding formula
for Yang-Mills,
one first includes in $S$ the action of the spin variables
(\ref{spinaction}) and
the dummy ghost variables (see \ref{dummyunity}). Next,
one must add terms in the exponent corresponding
to the second and third lines of (\ref{isingsum}) that 
rearrange the bond patterns of the new  spin and
ghost variables appropriately in the neighborhood of each solid line.
Finally, for  ${\cal V}_0$ and
${\bar{\cal V}}_0$ one substitutes the Yang-Mills 
Grassmann even versions 
of these quantities ${\cal V}_e$ and
${\bar{\cal V}}_e$ given in (\ref{evenvees}).

\begin{figure}[ht]
\begin{center}
$\epsfig{file=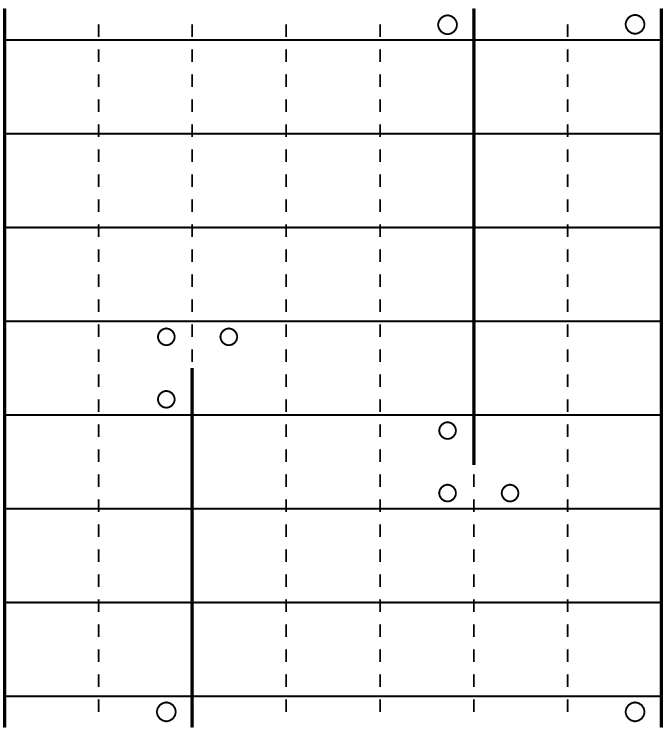,height=5cm}\hskip2cm
\epsfig{file=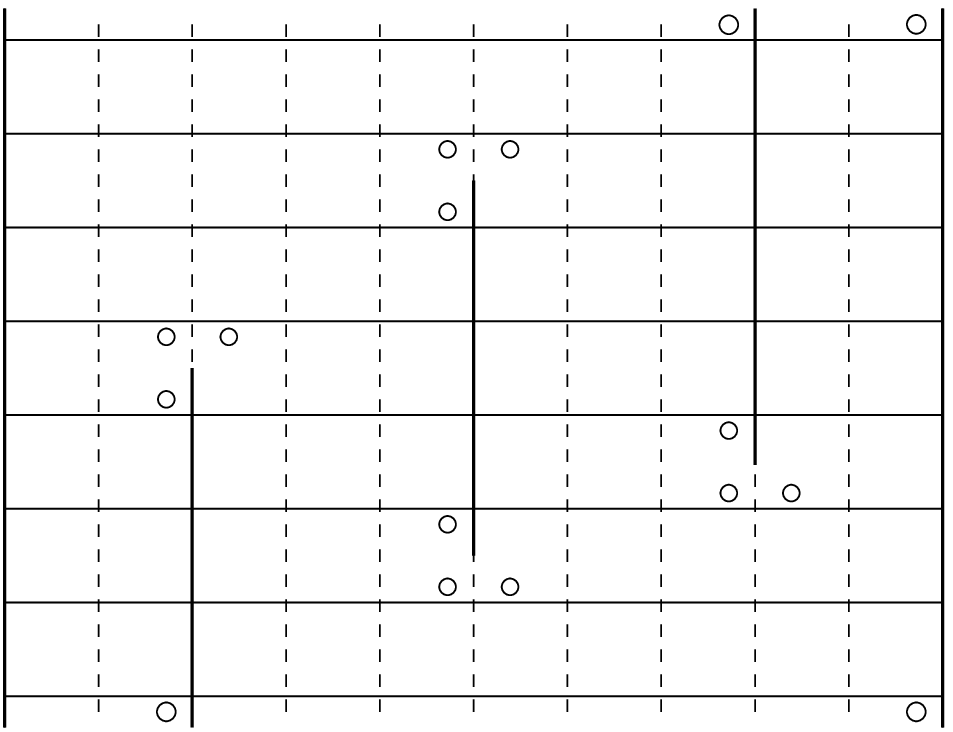,height=5cm}$
\end{center}
\caption{Four gluon diagram comprised of two cubics and
a one loop diagram. Only
the spin insertion locations are shown; those
in the bulk have been suppressed.}
\label{fourglueamp}
\end{figure}

As we have already discussed, the quartic
vertices of the Yang-Mills light-cone gauge
Feynman diagrams  are automatically produced by the fluctuation
contribution of two $\Delta{\bf q}$ insertions when
they occur on the same time slice and on the same
strip. Thus the formula for the sum of all planar diagrams
for Yang-Mills closely follows the scalar
$\Tr\phi^3$ paradigm. 
We draw a couple of four gluon diagrams in Fig.~\ref{fourglueamp},
and it should be evident how they are represented in
the worldsheet formalism. 

However, this is not the end of the story because 
we know that the continuum limit involves
divergences that require renormalization. It will
be very interesting to see how the renormalization
program can be carried out using our world sheet 
language. The $x^+,p^+$ lattice we have used
in our construction is identical to that
in Ref.~\cite{beringrt,gudmundssont} where
one loop diagrams are studied in some detail.
The worldsheet version of these calculations will
only differ in the treatment of end points of the
sums over discretized $x^+, p^+$ locations of the
vertices.
For example, the self-energy counter-term
required to ensure that the gluon stay massless
in perturbation theory has been explicitly
worked out in \cite{beringrt}. It is clear that
such counter-terms will have the structure of ``short loops'',
\ie\ vertical solid lines on the world sheet spanning
a small number of lattice sites. However,
we leave the analysis of renormalization and determination of
the necessary counter-terms in 
the worldsheet formalism for future work.

Another direction for future research is to repeat the
construction for supersymmetric field theories.
In particular, it will be very interesting to do
this for ${\cal N}=4$ supersymmetric Yang-Mills theory
in the limit $N_c\to\infty$.
This is the theory Maldacena
has conjectured is
dual to type IIB string theory on $AdS_5\times S_5$
\cite{maldacena,wittenholog,gubserkp}.
The worldsheet sigma model corresponding to this proposal
can be analyzed in the strong 't Hooft coupling limit $N_cg_s^2\to\infty$,
but it becomes a strong coupling theory in the opposite
limit $N_cg_s^2\to0$. Our construction, which
would have parallels to one previously proposed by Nastase and Siegel
\cite{nastasesiegel}, would give a two
dimensional worldsheet system that is tied to this
opposite limit. Maldacena duality should then mean that
our worldsheet system is dual to the $AdS_5\times S_5$
sigma model worldsheet system. Such a duality would
be fascinating to discover.

\vskip.25cm
\noindent\underline{Acknowledgments:}
I would like to thank K. Bardakci for the many
valuable insights he shared in our first collaboration,
and for his valuable comments on this work.
I thank J. Sonnenschein for useful discussions and
for reading the manuscript. Finally
I acknowledge useful discussions with D. Berenstein,
I. Klebanov, J. Maldacena, A. Polyakov and M. Strassler.
This work was supported in
part by the Monell Foundation and in part by
the Department of Energy under Grant No. DE-FG02-97ER-41029.

\bibliography{larefs,qcdsheetrefs}
\bibliographystyle{unsrt}

\end{document}